\documentclass[prb,preprint]{revtex4}

\usepackage{graphicx}
\usepackage{amsmath}
\begin{document}

\title{Pedagogical applications of the one-dimensional Schr\"odinger's equation to proximity effect systems: Comparison of Dirichlet and Neumann boundary conditions} 

\author{P. R. Broussard} 
\affiliation{Covenant College, Lookout Mountain, Georgia 30750}

\begin{abstract}

Proximity effect systems in superconducting films can be modeled by a one-dimensional Schr\"odinger equation. Several systems are studied using Dirichlet and Neumann boundary conditions. It is observed that the two boundary conditions have a dramatic effect on the lowest eigenstate allowed in these systems, and points to unusual behavior for solutions of Schr\"odinger's equation in certain potential wells and proximity effect systems.

\end{abstract}

\maketitle

\section{Introduction}
The superconducting proximity effect is the change in the superconducting transition temperature of a superconducting film when another metallic film with either a lower or zero superconducting transition temperature is deposited onto it, resulting in a proximity effect bilayer. There is much interest in these systems as transition edge calorimeters, such as the Mo/Cu bilayers being developed for detecting radiation.\cite{Rudman} Predicting the transition temperature of such systems is usually done with the de Gennes-Werthamer theory\cite{DeGennes, Werthamer} or the theory of Usadel.\cite{Usadel} These theories are difficult to introduce at the level of introductory quantum mechanics or solid state physics. However, it is possible to make a one-to-one correspondence between the problem of proximity effect in layered films and the quantum mechanics textbook example of the energy levels and wave functions of a particle in a one-dimensional potential well.\cite{Werthamer}

Reference~\onlinecite{Werthamer} derives the standard equation of proximity effect systems in the dirty limit (electron mean free path much less than the superconducting coherence length) given by (in one dimension)
\begin{equation}
\chi \Big(-\xi^{2}\frac{d^{2}}{d x^{2}}\Big)\Delta(x) +\ln \Big(\frac{\theta_{D}}{T_{c}(x)} \Big)\Delta(x)=\ln \Big(\frac{\theta_{D}}{T_{c}} \Big)\Delta(x),
\label{Werth}
\end{equation}
where $\chi(z) = \psi(1/2+z/2)-\psi(1/2)$, with $\psi(x)$ here being the diGamma function,\cite{AS} $\Delta(x)$ is the energy gap or self-energy\cite{Silvert} at the point $x$, $\xi$ depends on material properties in the superconductor or normal metal at point $x$, $\theta_{D}$ and $T_c(x)$ are the Debye temperature and the superconducting transition temperature for the material of interest at $x$ in isolation, and $T_{c}$ is the superconducting transition temperature of the composite structure.  Typically $T_{c}(x)$ is assumed to be uniform within a particular layer.  This equation is the differential equivalent of an integral equation for the proximity effect, and is usually much easier to solve.  Typically it is solved by expressing the self energy as a Fourier series and looking for the allowed wavevectors.  As shown in Reference~\onlinecite{Werthamer}, the above equation will predict the superconducting transition temperature of a proximity effect bilayer, however the solutions of this equation are difficult to present in undergraduate courses.

Reference~\onlinecite{Werthamer} discussed how equation~(\ref{Werth}) looks like the one-dimensional time independent Schr\"odinger equation, 
\begin{equation}
-\frac{\hbar^{2}}{2m}\frac{d^{2}\psi(x) }{d x^{2}}+ V(x)\psi(x)=E\psi(x)
\label{Sch}
\end{equation}
where $\psi(x)$ in this equation (and in the rest of the paper) is the wavefunction which is related to $\Delta(x)$, $\hbar^{2}/2m$ is related to $\xi^{2}$, $V(x)$ is related to $\ln(\theta_{D}/T_{c}(x))$, and $E$ is related to $\ln(\theta_{D}/T_{c})$. (The simplifying assumption is made that the quantity $\xi$ is the same for both layers, which in the Schr\"odinger equation is equivalent to assuming that the mass of the particle is the same in both layers.) Because solutions for constant potentials are easy to obtain with Eq.~(\ref{Sch}), it can be used to help students see how the transition temperature of proximity effect systems changes as layer thicknesses and potentials are modified. The purpose of this paper is to give some example solutions for various geometries and different boundary conditions and discuss how the solutions differ from what is typically expected.

\section{Description of the Specific Problem}

Although solutions to Eq.~(\ref{Sch}) will not give precise answers to the values of $T_{c}$ in a real system of metal layers, they will reproduce the overall behavior, provided we recognize several limitations. First, the energy eigenvalues of Eq.~(\ref{Sch}) are inversely related to the transition temperatures. The lowest energy eigenstate corresponds to the highest transition temperature. Because $\psi$ is related to the function $\Delta$ in Eq.~(\ref{Werth}), we also need to consider the boundary conditions. Although $\Delta$ and its derivative are not continuous,\cite{Werthamer} we will use the continuity of $\psi$ and $d\psi/dx$ to solve a simpler problem, and see that it still reproduces almost the same equation used in Ref.~\onlinecite{Werthamer}. (The actual boundary conditions used in solutions of the superconducting proximity effect are given in Ref.~\onlinecite{DeGennes}.) Another issue is what to do with the end surfaces, which in a proximity system are the metal/vacuum interfaces. In the usual particle in the box problem we exclude the possibility of the particle being outside the box, and we force $\psi$ to go to zero at a boundary. However, for $\Delta$ in layered superconductors, the correct boundary condition at a metal/vacuum boundaries is $d\Delta/dx=0$, which implies that the Cooper pairs are reflected off these boundaries back into the metal layer.\cite{DeGennes} We will compare similar geometries with two boundary conditions at the extreme boundaries, a Dirichlet type ($\psi=0$) and a Neumann type ($d\psi/dx=0$).

\begin{figure}[h!]
\begin{center}
\includegraphics*[width=6 in]{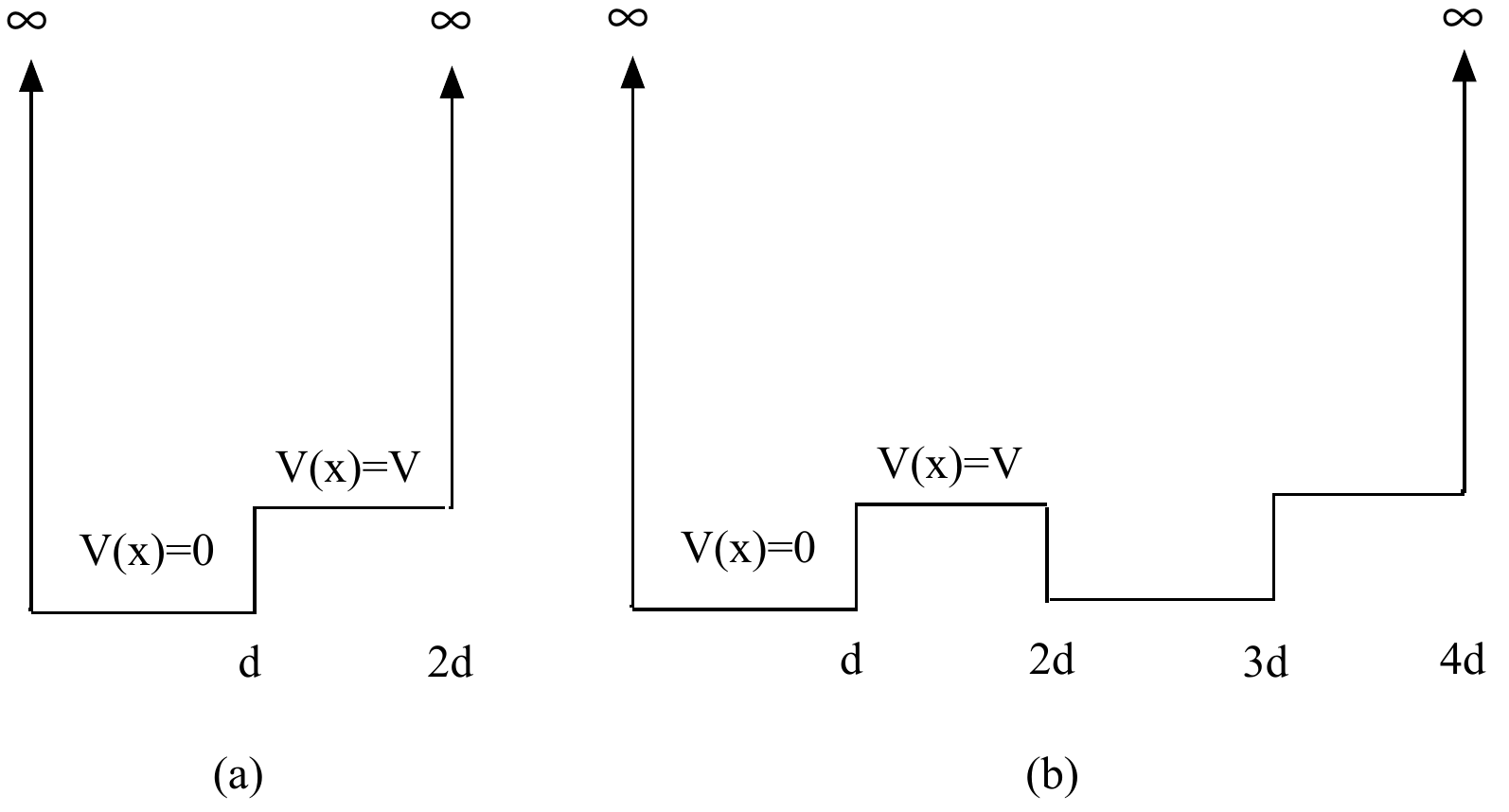}
\end{center}
\caption{Potential wells for bilayers, shown for (a) one period and (b) two periods. The widths of the individual layers is the same.}
\label{bilayer}
\end{figure}

To mimic a bilayer system the potential is as shown in Fig.~\ref{bilayer}(a), where the zero potential for $0<x<d$ represents a material with a high transition temperature, and the finite value of $V$ for $d<x<2d$ represents a material with a lower transition temperature. (We can set the lower potential to any value we choose to adjust the zero of the energy.) Because the probability density is assumed to vanish outside these two layers, there are infinite barriers at the edges. Any sequence of high and low transition temperature layers can be mimicked, as seen in Fig.~\ref{bilayer}(b) for a 2 period proximity system, and it would be easy to vary the ratio of the thicknesses of the high and low potential. For the purpose of this paper equal thickness layers will be used.

Rather than changing the thickness of the layers, the potential $V$ will be varied and eigenvalues and eigenstates found. There is a one-to-one correspondence between the two, for in our solutions all energies and potentials will be given in units of $\hbar^{2}/(2md^{2})$. For example for a bilayer with width $2d$, the lowest energy eigenstate would be $\pi^{2}/4$ with $V=0$, the lowest one for an infinite square well of width $2d$ (with a Dirichlet boundary condition). So changing $V$ is equivalent to changing $d$ and vice versa.

We have to interpret the meaning of $\psi$ for the case of the proximity effect. As stated in Ref.~\onlinecite{DeGennes} when $\Delta(\mathbf{r})$ depends on only one space coordinate, $x$, the energy gap $E_{0}$ is equal to the minimum value of $|\Delta(x)|$ in the sample. This role of $\Delta$ means that it can not be allowed to go to zero or be negative,\cite{foot} and hence $\psi$ must be positive definite. In addition, the transition temperature of a bilayer system will always be between the transition temperatures of the two materials so that only solutions with energy eigenvalues satisfying $0<E<V$ will be studied. There are valid solutions with $E>V$, but they will not shed light on the proximity effect.

\section{One Period}

We first look at a simple bilayer, as seen in Fig.~\ref{bilayer}(a), which in the absence of the potential $V$ would have the lowest eigenstate $E=\pi^2/4$. We seek a solution that satisfies $0<E<V$. Because the potentials are constant and $0<E<V$, we know $\psi$ can be written in the low potential layer (layer 1 with the higher transition temperature) as $\psi_{1}(x)=\alpha_{1} \cos(kx) +\beta_{1} \sin(kx)$ where $k=\sqrt{2mE/\hbar^{2}}$; in the high potential layer (layer 2 with the lower transition temperature) $\psi_{2}(x)=\alpha_{2} \cosh(qx) +\beta_{2} \sinh(qx)$ where $q=\sqrt{2m(V-E)/\hbar^{2}}$. The continuity of $\psi$ and $d\psi/dx$ will be enforced at $x=d$, but we first have to consider the boundaries at the extreme edges, which represent the metal/vacuum interface. 

\begin{figure}[h]
\begin{center}
\includegraphics*[width=4.5in]{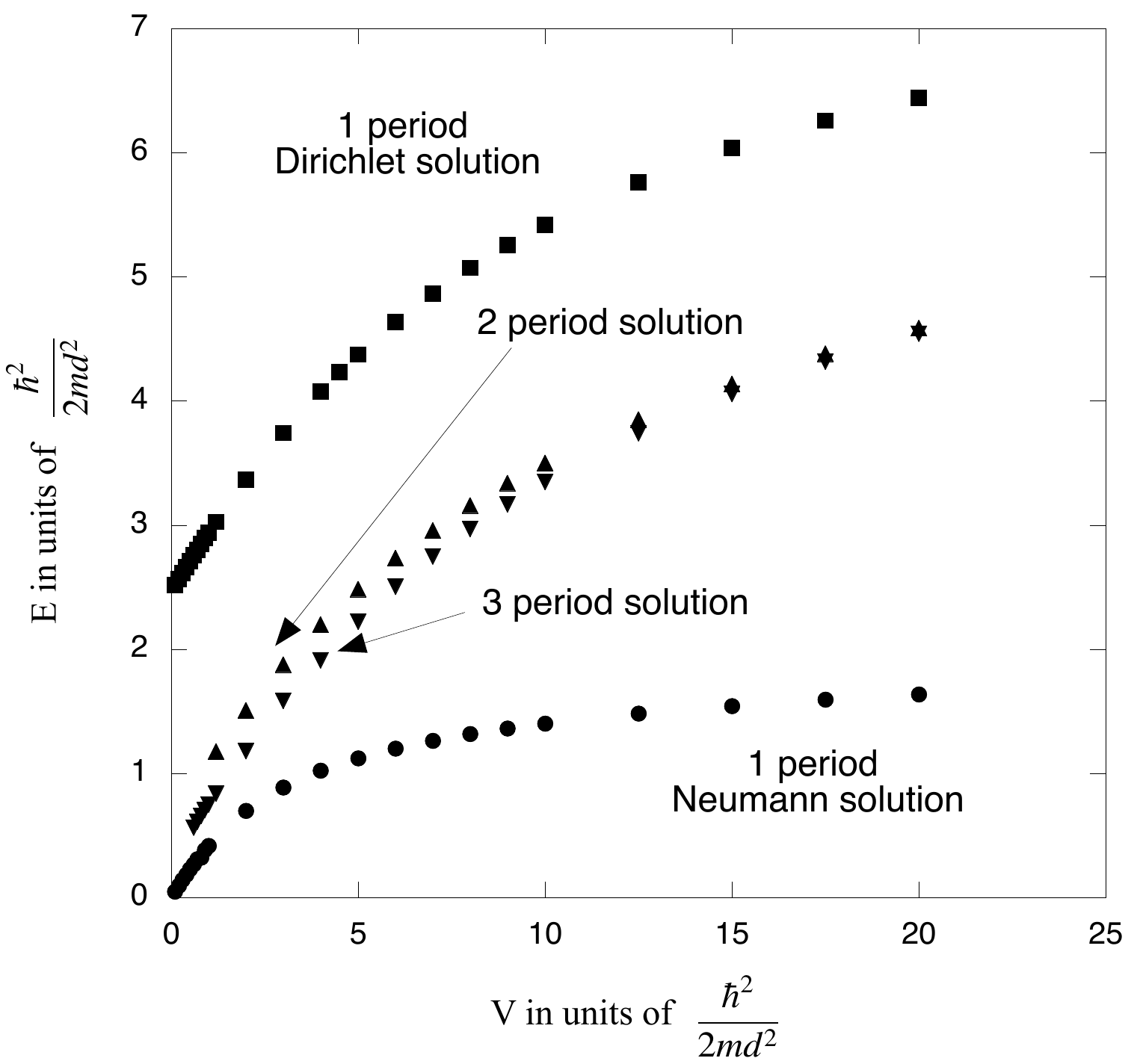}
\end{center}
\caption{Energy eigenvalues versus potential step height for the one period Dirichlet and Neumann boundary condition solutions, Eqs.~(\ref{1D}) and (\ref{1N}), respectively; the two period solution with the one period solution removed, Eq.~(\ref{2Dr}); and the three period solution with the one period solution removed, Eq.~(\ref{3Dr}). Notice the Neumann eigenvalues for one period are always below those for the Dirichlet solution for one period or the common solutions for multiple periods.}
\label{Eng}
\end{figure}

We first look at the Dirichlet boundary condition, which forces $\psi$ to be zero at $x=0$ and $x=2d$. For this case the first layer will have $\psi_{1}(x)=\beta_{1} \sin(kx)$ and for the second layer $\psi_{2}(x)=\beta_{2} \sinh [q(2d-x)]$. Continuity of $\psi$ at $x=d$ gives $\beta_{1}\sin(kd)=\beta_{2} \sinh(qd)$, and continuity of the derivative of $\psi$ gives $k\beta_{1}\cos(kd)=-q\beta_{2} \cosh(qd)$. The ratio of these two conditions gives,
\begin{equation}
q\tan(kd)+k\tanh(qd)=0.
\label{1D}
\end{equation}
The roots of Eq.~\eqref{1D} and all others were found graphically;\cite{grapher} recall that the values of $E$ and $V$ are given in units of $\hbar^{2}/2md^{2}$. For this case there are no eigenstates with energies less than $V$ until $V\gtrsim4.12$; the energy eigenvalues are shown in Fig.~\ref{Eng}. There is a lower energy eigenstate for $V<4.12$, but it is greater than $V$. (For $V < 4.12$ the lowest solutions are given by solving $q\tan(kd)+k\tan(qd)=0,$ whose solutions are plotted in Fig.~\ref{Eng}.) The limit $V\rightarrow\infty$ is equivalent to an infinite well with width 1/2, or $E=\pi^2\approx 9.870$.

For the Neumann condition the wavefunction must have a zero slope at $x=0$ and $2d$. The form that satisfies these conditions is $\psi_{1}(x)=\alpha_{1} \cos(kx)$ in the first layer and $\psi_{2}(x)=\alpha_{2} \cosh(q(2d-x))$ in the second layer.  The continuity of $\psi$ and its derivative gives $\alpha_{1}\cos(kd)=\alpha_{2}\cosh(qd)$ and $-k\alpha_{1}\sin(kd)=-q\alpha_{2}\sinh(qd)$. The ratio gives the energy eigenvalue equation,
\begin{equation}
k\tan(kd)=q\tanh(qd),
\label{1N}
\end{equation}
which is the same as Eq.~(14) in Ref.~\onlinecite{Werthamer} and similar to Eq.~(4.16) in Ref.~\onlinecite{DeGennes}. In contrast to the eigenvalues for the Dirichlet boundary condition, the energy eigenvalues satisfy $E<V$ for all values of $V$. If $V=0$, which would represent a layer of material with the higher transition temperature with width $2d$, $T_{c}$ of the bilayer would be the same as that of the higher transition temperature layer, which we represent by $E=0$. The energy eigenvalues are shown in Fig.~\ref{Eng} in comparison to those for the Dirichlet condition. We see, consistent with the exact proximity effect equations, that as $d$ increases (or $V$ decreases), $E$ approaches zero, and the predicted transition temperature approach the higher $T_{c}$ material; as $d$ decreases, the predicted transition temperature decreases.

\begin{figure}[h!]
\begin{center}
\includegraphics*[width=4.5in]{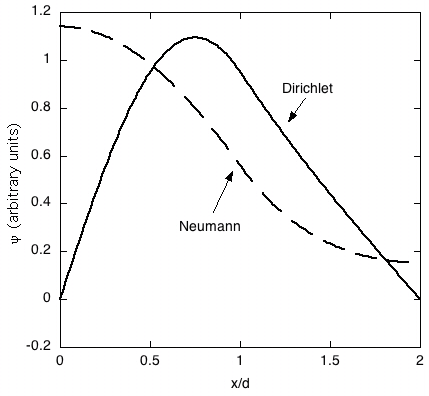}
\end{center}
\caption{Wavefunctions versus $x/d$ for the one period system for Dirichlet and Neumann boundary conditions. Here $V=5$. The energy eigenvalues are $E=$ 4.38 for Dirichlet case and 1.12 for the Neumann case.}
\label{1Dplot}
\end{figure}

The corresponding wavefunctions are compared in Fig.~\ref{1Dplot} for $V=5$. As expected the Dirichlet condition puts the bulk of the probability in the lower well. In the Neumann case the shape is what is expected for a standard bilayer proximity effect system, with the superconducting order parameter decaying into the non-superconducting system, as shown in Ref.~\onlinecite{DeGennes}. The solutions in Fig.~\ref{1Dplot} have been properly normalized, but for the proximity effect (Neumann) case there is no clear normalization condition, because the minimum of the gap function would be the measured energy gap in the system.\cite{DeGennes}

\section{Two Periods}

Next we turn to the effect multiple layers will have on the lowest eigenvalue of our quantum potential, or the transition temperature of a multiple layered structure.  Consider the potential is shown in Fig.~\ref{bilayer}(b), which would represent a two period bilayer.  For Dirichlet boundary conditions $\psi$ must be chosen in the first layer as $\psi_{1}(x)=\beta_{1}\sin(kx)$ and in the last layer as $\psi_{4}(x)=\beta_{4}\sinh[q(4x-d)]$. For the interior layers (2 and 3) it is convenient to chose $\psi_{2}(x)=\alpha_{2}\cosh[q(2d-x)]+\beta_{2}\sinh[q(2d-x)]$ and $\psi_{3}(x)=\alpha_{3}\cos[k(x-2d)]+\beta_{3}\sin[k(x-2d)]$. Continuity of $\psi$ and its derivative at the three interior boundaries gives six relations, which can be reduced to eliminate the coefficients in $\psi$ giving the equation for the energy eigenvalues,
\begin{equation}
[2kq+(q^{2}-k^{2})\tan(kd)\tanh(qd)] u_{D}=0,
\label{full2D}
\end{equation}
where $u_{D}=[q\tan(kd)+k\tanh(qd)]$ is the solution of the one period Dirichlet condition problem in Eq.~(\ref{1D}). Thus, solutions to the one period case are also solutions to the two period case.

\begin{figure}[h!]
\begin{center}
\includegraphics*[width=4.5in]{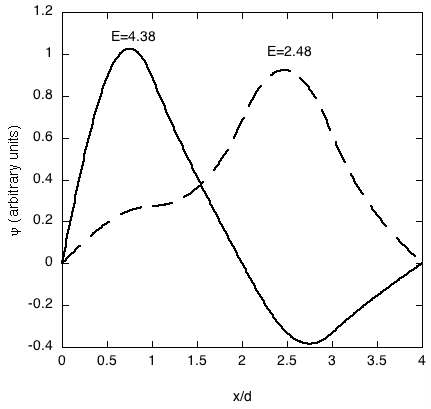}
\end{center}
\caption{Comparison of wavefunctions for the two period case (Dirichlet condition) with $V=5$ for $E=4.38$ (one period solution) and $E=2.48$ given by Eq.~\ref{2Dr}).  The one period solution puts the bulk of the probability in the first layer, while the two period solution puts the bulk in the third layer, allowing for a lower energy eigenvalue. }
\label{2Dplot}
\end{figure}

For the moment we will assume that $u_{D} \neq 0$, so that the two period eigenvalue equation becomes
\begin{equation}
2kq+(q^{2}-k^{2})\tan(kd)\tanh(qd)=0.
\label{2Dr}
\end{equation}
The two period energies given by removing the one period solutions are shown in Fig.~\ref{Eng} and are lower than those of the one period Dirichlet solution. In Fig.~\ref{2Dplot} the two wavefunctions for the two period case are shown, one using the energy eigenvalue of the one period case, Eq.~(\ref{1D}), and the other the energy eigenvalue for the two period case with the one period solution removed, Eq.~(\ref{2Dr}). In both cases $V=5$. The bulk of the probability for the lower energy state is in the third layer. For the higher energy state the bulk is in the first layer, where the Dirichlet condition in combination with the one period solution (which creates a node at $x/d=2$) forces a smaller wavelength, resulting in a larger energy. The result is what we would expect, namely, if the system size increases, the energy eigenvalues of the system decrease. The interesting feature is that there is a lower energy eigenvalue that appears when the system goes to two periods, but the original energy eigenvalue does not change as the system size increases.

\begin{figure}[h!]
\begin{center}
\includegraphics*[width=4.5in]{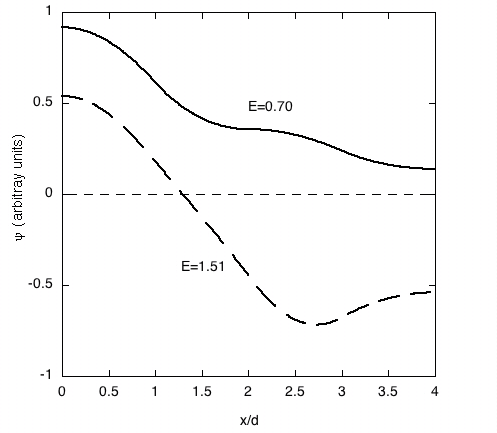}
\end{center}
\caption{Comparison of wavefunctions for the two period case (Neumann condition) with $V=2$ for $E=0.70$ (one period solution) and $E=1.51$ given by Eq.~(\ref{2Dr}).  Here the two period solution forces the wavefunction to be negative over most of $x/d$.}
\label{2Nplot}
\end{figure}

For the Neumann condition the only modification for two periods is that the values of $\psi_1(x)$ and $\psi_4(x)$ are changed to $\alpha_{1}\cos(kx)$ and $\alpha_{4}\cosh[q(4x-d)]$, respectively, to enforce zero slope for $\psi$ at the edges. This change and continuity of $\psi$ and its derivative in the interior three boundaries give the energy eigenvalue equation,
\begin{equation}
[2kq+(q^{2}-k^{2})\tan(kd)\tanh(qd)]u_{N}=0,
\label{2Nfull}
\end{equation}
where $u_{N}=[q\tanh(qd)-k\tan(kd)]$, which is the same as Eq.~(\ref{1N}). Equation~\eqref{2Nfull} is almost the same as derived in Ref.~\onlinecite{Broussard} for the full DeGennes-Werthamer solution for a two period bilayer; the interest here is how the eigenvalues and eigenfunctions compare. As was seen for Dirichlet boundary conditions, the one period solution is also a solution to the two period case, so the repetition of the single period solution is not due to the boundary condition. As before, $u_{N}$ can be factored out to give the reduced eigenvalue equation which is the same as that for the Dirichlet boundary condition, Eq.~(\ref{2Dr}). Comparison of these energy values to the one period Neumann condition is also shown in Fig.~\ref{Eng}, but here the energies for Eq.~(\ref{2Dr}) are higher than those for the one period Neumann condition, Eq.~(\ref{1N}). Thus, in the Neumann case, increasing the size of the system does not lower the lowest energy eigenvalue. As can be seen in Fig.~\ref{2Nplot}, the one period solution enforces a zero slope at the $x=2d$ boundary, and creates a reduced version of the first half for the second half. The other solution has a node in the wavefunction which is not allowed for $\Delta(x)$ for the proximity effect.

For the two period case there is an allowed solution which repeats the one period solution, but whose energy compares with the lowest eigenvalue very differently. For the Dirichlet condition it is higher, and for the Neumann condition it is lower than the common energy eigenstate, whose energy does not depend on the boundary conditions. Thus, although there are energy eigenstates that are similar for the two boundary conditions, there are also differences that can be used to show the importance of boundary effects.

\section{Three Periods}

\begin{figure}[h!]
\begin{center}
\includegraphics*[width=4.5in]{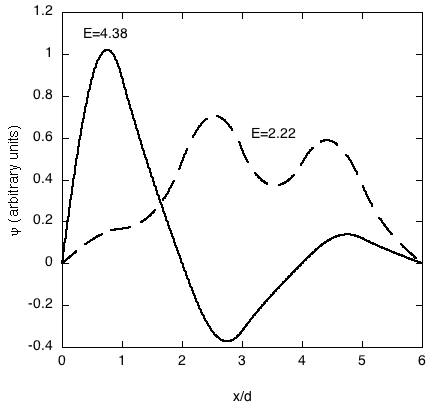}
\end{center}
\caption{Comparison of wavefunctions for the three period case (Dirichlet condition) with $V=5$ for $E=4.38$ (one period solution, Eq.~(\ref{1D})) and $E=2.22$ (three period solution, Eq.~(\ref{3Dr})).  Similar to the two period case, the three period solution has the bulk of the probability in the third and fifth layers.}
\label{3Dplot}
\end{figure}

In this case the potential has three repeating sections going from $x/d =0$ to $6$, and the complexity of the calculations increases substantially. We choose Dirichlet conditions on the ends and label the layers as 1,\dots, 6. We must choose $\psi_{1}(x)$ as before, but choose $\psi_{6}(x)=\beta_{6}\sinh[q(6d-x)]$. We use the same values of $\psi_{2}(x)$ and $\psi_{3}(x)$, with $\psi_{4}(x)=\alpha_4\cosh[q(4d-x)]+\beta_4\sinh[q(4d-x)]$ and $\psi_{5}(x)= \alpha_5\cos[k(x-4d)]+ \beta_5 \sin[k(x-4d)]$.  Continuity of $\psi$ and its derivative at the five interior boundaries gives ten equations, which after much algebra can be reduced to an equation for the energy eigenvalue:
\begin{align}
u_{D}\Big\{(k^{4}-k^{2}q^{2}+q^{4})\tan^{2}(kd)\tanh^{2}(qd)
+ k^{2}q^{2}(3+\tanh^{2}(qd)\nonumber\\-\tan^{2}(kd))+4kq(q^{2}-k^{2})\tan(kd)\tanh(qd)\Big\}=0.
\label{3D}
\end{align}
we again factor out the one period solution giving, 
\begin{align}
(k^{4}-k^{2}q^{2}+q^{4})\tan^{2}(kd)\tanh^{2}(qd) &+k^{2}q^{2}(3+\tanh^{2}(qd)-\tan^{2}(kd))\nonumber \\
& \quad +4kq(q^{2}-k^{2})\tan(kd)\tanh(qd)=0.
\label{3Dr}
\end{align}
As expected from the two period case, Eq.~\eqref{3Dr} has slightly lower energy eigenvalues than the two period case, Eq.~(\ref{2Dr}), and there are valid solutions for lower values of $V$ than the two period case, as seen in Fig.~\ref{Eng} for the points labeled as three period solution. These solutions are lower in energy than the one period Dirichlet solutions, which can be understood by looking at the wavefunctions in Fig.~\ref{3Dplot}. The solution for the one period eigenvalue shows how the wavefunction is replicated and reduced as $x/d$ increases. The solution for the three period equation, Eq.~(\ref{3Dr}), puts the majority of the probability in the second and third wells, in contrast to the one period solution which has most of the probability in the first well. Thus, as in the two period case, the lowest energy eigenvalue is lower, but there remains a state whose energy eigenvalue remains unchanged as the number of periods increases.

\begin{figure}[h!]
\begin{center}
\includegraphics*[width=4.5in]{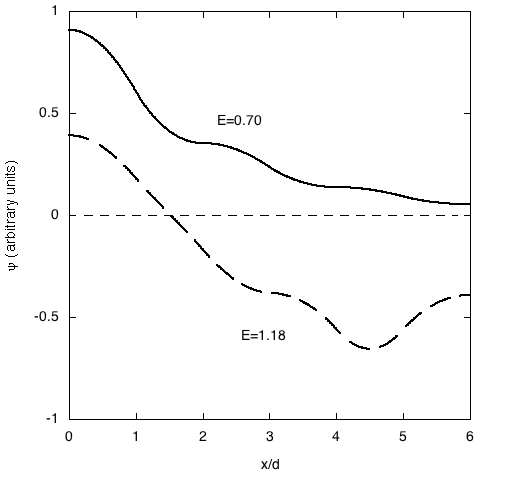}
\end{center}
\caption{Comparison of wavefunctions for the three period case (Neumann condition) with $V=2$ for $E=0.70$ (one period solution, Eq.~(\ref{1N})) and $E=1.18$ (three period solution, Eq.~(\ref{3Dr})).  As in the two period case, the three period solution is negative over most of $x/d$ again.}
\label{3Nplot}
\end{figure}

For the Neumann boundary condition the wavefunctions for the first and last layer are changed to $\psi_{1}(x)=\alpha_{1}\cos(kx)$ and $\psi_{6}=\alpha_{6}\cosh[q(6d-x)]$, with all the others from the Dirichlet condition unchanged. The continuity of the wavefunction and its derivative can be used to produce ten equations which can be reduced to the following equation for the energy eigenvalue,
\begin{eqnarray}
u_{N}\Big\{(k^{4}-k^{2}q^{2}+q^{4})\tan^{2}(kd)\tanh^{2}(qd)+k^{2}q^{2}(3+\tanh^{2}(qd)-\tan^{2}(kd))\nonumber \\
{}+4kq(q^{2}-k^{2})\tan(kd)\tanh(qd)\Big\}=0.
\end{eqnarray}
The solution consists of the product of the one period Neumann condition, along with the same equation as in Eq.~(\ref{3Dr}). As for the two period Neumann solution, the energy states are the one period solution (which is the lowest) followed by the same energies as the three period Dirichlet condition. Comparison of the wavefunctions shown in Fig.~\ref{3Nplot} shows the one period solution, with the amplitude decreasing as the power of the number of periods; the three period solution has a node and negative values over much of the range of $x/d$, which would be unrealistic for the energy gap function $\Delta$.

From the structure of these solutions it is apparent that for a $N$ period solution, the Dirichlet and Neumann boundary conditions have an energy eigenvalue that is independent of $N$, which is the lowest energy eigenstate for the Neumann case, although a proof of this property has not been shown here. In addition, these results imply there is a common energy eigenstate between the two conditions, which is the lowest energy eigenstate for the Dirichlet case.

\section{Application to Proximity Effect Systems}

In this study, the idea that the lowest energy eigenvalue should decrease as the system size increases is  seen to be strongly dependent on the boundary conditions used.  For Dirichlet conditions the decrease is observed, while for Neumann conditions, the lowest energy eigenvalue is unchanged as the system size increases.  As stated in Ref.~\onlinecite{Werthamer} the exact proximity equations are similar to a one-dimensional Schr\"odinger's equation with simple potentials and Neumann conditions. The one period system faithfully reproduces the behavior of a simple bilayer. The results from this study imply that the transition temperature for a $N$ period bilayer system (which correlates to the lowest energy eigenvalue) would be independent of $N$.\cite{Broussard} (In Ref.~\onlinecite{Broussard} one-period, two-period and three-period structures were measured.) However the shape of the resulting wavefunction found here is unexpected. If the minimum value of $\Delta(x)$ is the energy gap in the system,\cite{DeGennes} then as the number of periods increases, either the gap would become smaller, or the maximum of the self-energy function would increase for a fixed gap, neither of which is reasonable, given that the transition temperature remains the same. The actual behavior of such a system is still an open question.   The simplicity of this approach would allow other choices of potential energy well structures to be studied in order for students to model various layered films, such as trilayers, or fractal like layer arrangements.

\begin{acknowledgments}
The author would like to express his appreciation to the two referees who made valuable suggestions which improved the readability of the paper substantially.
\end{acknowledgments}

\newpage
\section{Figures}

\begin{thebibliography}{}

\bibitem{Rudman}
K. D. Irwin, G. C. Hilton, John M. Martinis, S. Deiker, N. Bergren, S. W. Nam, D. A. Rudman, and D. A. Wollman, ``A Mo-Cu superconducting transition-edge microcalorimeter with 4.5\, eV energy resolution at 6\,keV," Nucl. Instr. Meth. A {\bf 444}, 184--187 (2000).

\bibitem{DeGennes}P. G. de Gennes,
``Impurity and boundary effects in superconductors," Rev. Mod. Phys. {\bf 36}, 225--237 (1964).

\bibitem{Werthamer}N. R. Werthamer,
``Theory of the superconducting transition temperature and energy gap function of superposed metal films," Phys. Rev. {\bf 132}, 2440--2445 (1963).

\bibitem{Usadel}Klaus D. Usadel,
``Generalized diffusion equations for superconducting alloys,'' Phys. Rev. 
Lett. {\bf 25}, 507--509 (1970). 

\bibitem{AS} Milton Abromowitz and Irene A. Stegun, {\it Handbook of Mathematical Functions} (Dover, New York, NY, 1972),  pp.\ 258-259.

\bibitem{Silvert}William Silvert,
``Theory of the superconducting proximity effect," J. Low Temp. Phys. {\bf 20}, 439--477 (1975).

\bibitem{Broussard}P. R. Broussard,
``Boundary-condition effects on the superconducting transition temperature of proximity-effect systems," Phys. Rev. B {\bf 43}, 2783--2787 (1991).

\bibitem{foot}Reference~\onlinecite{DeGennes} uses the absolute value of $\Delta$ because materials with a repulsive electron interaction have negative values of $\Delta$. In this paper $V$ is assumed to be positive definite.

\bibitem{grapher} We used the program Grapher on Mac OSX.

\end{thebibliography}
\end{document}